\documentstyle[12pt,aaspp]{article}
\def\um{{\,\mu\rm m}}

\def\cline{\noalign{\vskip 16pt\hrule \vbox to 3pt{}}}

\def\mag{\,{\rm mag}}
\def\@journalname{Astronomical Journal}
\def\accepted#1{\gdef\@accptdate{#1}} \accepted{\relax}
\def\journalid#1#2{\gdef\@jourvol{#1}\gdef\@jourdate{#2}}
\def\articleid#1#2{\gdef\@startpage{#1}\gdef\@finishpage{#2}}
\begin{document}
\title{Infrared Counterpart of the Gravitational Lens 1938+66.6}

\author{James E. Rhoads, Sangeeta Malhotra, \& Tomislav Kundi\'{c} }
\affil{Princeton University Observatory, Peyton Hall, Princeton, NJ
08544 \\
rhoads@astro.princeton.edu,san@astro.princeton.edu,
tomislav@astro.princeton.edu}  

\begin{abstract}

We report the detection of a very red source coincident with the
gravitational lens 1938+66.6 (Patnaik et al.  1992) in $K'$ ($2.12
\um$), $H$ ($1.6 \um$), $J$ ($1.25 \um$), and Thuan-Gunn $r$ ($0.65
\um$) bands.  1938+66.6 has previously been detected as a partial radio
ring indicating lensing.  We find  $K^\prime=17.1 \pm 0.1$ and $r = 23.9
\pm 0.2$, making it a very red source with $(r-K^\prime)=6.8 \pm 0.25$. 
We also observed in Thuan-Gunn $g$ band ($0.49 \um$) and found $g>24.5$
at the 90\% confidence level.  We interpret our observations as a reddened
gravitational lens on the basis of its optical-IR color and positional
coincidence with the radio source.

\end{abstract}

\section{Introduction} 

There is increasing evidence that quasars and gravitationally lensed
quasars sometimes have very red optical-infrared colors.  Reddening by
intervening dust offers an explanation for this phenomenon (Webster et
al. 1995, Lawrence et al. 1995, Larkin et al. 1994).  Such large
reddenings raise questions about possible incompleteness of optical
quasar and lens surveys.  This in turn can have important implications
for cosmological models, since many constraints on cosmological
parameters are derived from the luminosity functions of quasars and
from the numbers and seperations of lensed quasars (Turner, Ostriker
\& Gott 84, Fukugita and Peebles 1993, Fukugita \& Turner 1991, Malhotra
\& Turner 1995, Kochanek 1993, Maoz \& Rix 1993). The reddening of
gravitationally lensed quasars is also a potential probe of the ISM of
the lensing galaxies at high redshift.

One would expect the radio selected gravitational lenses to be free of
bias against reddened lenses. Also small seperation lenses, where the
optical paths are likely to pass near the center of a single lensing
galaxy, are more likely to show substantial amount of reddening.

The radio source 1938+66.6 is one of the small seperation
gravitational lenses discovered in the VLA/MERLIN survey of flat
spectrum radio sources (Patnaik et al 1992).  It has four compact
sources and an arc; the maximum seperation between the any of the
components is $0''.95$.  Patnaik et al report a r=23 object at the
location of the lens.  We observed this source in the near infrared
$K^{\prime}$ ($2.12 \um$), $H$ ($1.6 \um$), and $J$ ($1.25 \um$)
bands, and in the optical Thuan-Gunn $r$ ($0.65 \um$) and $g$ ($0.49
\um$) bands.  An object was detected at the position of the lens in
all bands except $g$.  This optical-infrared object will hereafter be
referred to as ``1938+666(OIR)''; we will argue that it is a
counterpart of the radio system.

The observations and data analysis are described in section 2. In
section 3 we estimate the probability of the infrared source being a
cool star or a high redshift galaxy on the basis of color. Discussion
and posssible interpretation of the results are presented in section
4.



\section{Observations}

We observed 1938+666 in $H$ band on 2 July 1995 (UT), in $J$ and $K'$
bands on 6 August 1995 (UT), and in Thuan-Gunn $g$ and $r$ bands on 4
and 24 July and 5 August 1995 (UT).  There was intermittent
scattered cloud cover early on the night of 2 July, but the sky
cleared shortly before the 1938+666 $H$ band observations.  All data
were taken with the Apache Point Observatory%
\footnote{APO is privately
owned and operated by the Astrophysical Research Consortium (ARC),
consisting of the University of Chicago, Institute for Advanced Study,
Johns Hopkins University, New Mexico State University, Princeton University,
University of Washington, and Washington State University.}
3.5 meter telescope.  The
near-IR observations used the GRIM II camera, while the $g$ and $r$
band observations used the Double Imaging Spectrograph (DIS) in
imaging mode.  Seeing was about $1.6''$--$1.8''$ FWHM in all bands.
Some parameters of the observations are summarized in table~1.

All data reduction was done using the IRAF package and followed
standard procedures.  For the near-IR data, individual short exposures
were sky subtracted, typically using the 2--4 sky frames nearest in
time.  A sky flat was made by taking the median of all disregistered
exposures in a band, subtracting a mean bias frame, and normalizing
the result by its median pixel value.  The sky subtracted image frames
were then divided by this flat.  Finally, the images were registered
(using the centroid of a bright star to determine fractional pixel
shifts) and a combined image was produced by taking the median of all
processed single frames at each pixel.  For the $g$ and $r$ bands
processing was similar but no sky subtraction step was needed.
Final images are shown in figure~1.


On the nights 2 July, 4 July, and 6 August we also observed
photometric standards to calibrate our fluxes.  For the $H$ band, we
used the UKIRT faint standard FS 28; for $J$ and $K'$ bands
we used FS 7 (Casali \& Hawarden 1992).  For the $g$ and $r$ bands, we
used the fields Mark A (Jorgensen 1994, Landolt 1992) and F1038-6
(Jorgensen 1994, Stobie et al 1985).


\subsection{Data Analysis}
We determined the position of the radio source using three HST
guide stars that fell within the r band image (epoch 2000.0
coordinates 19 38 14.23 +66 50 23.99, 19 38 45.36 66 48 40.38, and
19 38 01.38 +66 51 14.11).  We first used the guide stars to determine
the position and orientation of the coordinate grid,  and then
determined the position of the quasar relative to each guide star.
The scatter in the resulting position estimates was $0.7''$ in RA and
$0.4''$ in declination.



In the $r$, $J$, $H$, \& $K'$ bands, the counterpart to 1938+666 is
apparent, and we have done conventional aperture photometry to measure
its fluxes.  In the $g$ and $r$ bands we have applied additional tests
to see if an object is detected and to place upper limits on the
counterpart flux.  Our results are reported in Table~1.

We measured aperture magnitudes with radii $\sim 0.67 a$ where $a$ is
the FWHM (full width at half maximum) of the psf (point spread
function) and the coefficient $0.67$ maximizes the signal to noise for
aperture photometry of a faint source with a Gaussian psf.
%

To see if the resulting $g$ and $r$ fluxes are significantly above
background, we measure fluxes in randomly placed apertures and
determine how frequently the flux exceeds that seen for 1938+666.  The
result is not very sensitive to aperture size for $1.5'' \la a \la
2.0''$.  In $g$ band, $89\%$ of randomly placed $1.5''$ apertures had
fluxes exceeding that for 1938+666, and we conclude there is no
evidence of a $g$ band counterpart to the radio source.  In $r$ band,
$18\%$ of randomly placed $1.5''$ apertures had fluxes exceeding that
for 1938+666.  Visual inspection of the final $r$ band image shows a
source at the location of the NIR source, and we believe that the weak
result of the random aperture test is due to bleeding columns from
saturated bright stars and to the substantial density of 23rd
magnitude objects on the sky.

By modelling the counts in an aperture of area $A$ pixels as the
object flux plus sky noise, we may place an upper limit on the total
flux for a point source (modelled as a Gaussian psf) at the H band
source location.  At confidence levels of $(90\%, 99\%,$ and $99.9\%)$
we find $g > (24.5, 24.0,$ and $23.7) \mag$.  Applying the same formalism
to the weakly detected $r$ band source gives $r > (23.75, 23.6,$
and $23.5) \mag$.


A source substantially larger than the psf could be brighter than the
total magnitude limits quoted above.  Additionally, the spectral slope
of the putative candidate will influence the $r$ band limit, since our
standard star observations suggested a substantial color term ($-0.15
(g-r)$) in the transformation from instrumental to standard $r$
magnitude. No important color term was apparent in $g$ band.  We were
not able to observe enough NIR standards to allow secure color term
corrections in $J$, $H$, and $K'$ bands.

\section{Interpretation}

To test the interpretation of 1938+666(OIR) as the counterpart of the
radio source 1938+666, we compared its measured colors to those of
galactic stars, normal galaxies, quasars, and lensed systems.  We
first account for Galactic foreground dust.  Reddening towards
1938+666 can be estimated at $0.10 \le E(B-V) \le 0.12 \mag$,
based on the reddening map of Burstein \& Heiles (1982).  The normal
galactic extinction law (with $R_V \equiv A(V)/E(B-V) = 3.1$) yields
$E(r-K) \approx 2.25 E(B-V)$ (cf. Mathis 1990).  Thus, Galactic dust
does not contribute strongly to the extremely red color of
1938+666(OIR).

Our data indicate $(r-H) = 6.3 \pm 0.25 \mag$, $(r-K') = 6.8 \pm 0.25 \mag$,
$(J-H) = 2.1$, and $(H-K') = 0.5$.

For comparison, data in Bessell (1991) plus the color transformations
of Jorgensen (1994) give $ (r-H) \approx 6$ for the M6.5 dwarf LHS
3003, and $ (r-H) \approx 6.8$ for an M7.5 dwarf.  Thus, only the very
coolest stars have colors as red as 1938+666.

Two further tests may be applied to the hypothesis that the object is
a very cool star.  First, we can see if 1938+666(OIR) is appreciably
broader than a point source.  In the present $H$ band data, the FWHM
of 1938+666(OIR) is $\sim 2.4''$, while a representative stellar FWHM
is $\sim 1.8''$.  This would be consistent with a source of size $\sim
1''$, but better seeing and a larger signal to noise ratio are
required before this test can be considered conclusive.  Second, we
can use Wainscoat et al's (1992) model of the near-IR sky to calculate the
number density of ``M late V'' dwarfs with $18 < H < 19$ towards
1938+666; the resulting density is 0.25/arcmin$^2$, giving a
probability $\sim 2 \times 10^{-3}$ of finding a suitably red M dwarf
within 3 arcseconds of the radio source.

We can also compare the observed color limit to the colors of normal
galaxies.  The reddest local elliptical galaxies have $(r - H)$ colors
around $2.75$ (Persson, Frogel, \& Aaronson 1979), while moderate
redshift ellipticals ($z$ up to $0.92$) show $(r-H) \le 4$ (Persson 1988).
By redshift $z \approx 1.2$, elliptical galaxies can have $R-K$ up to $6$
(Dickinson 1995), and the central component of the $z = 2.016$ radio galaxy
0156$-$252 has $r - K = 7.3$ (McCarthy, Persson, \& West 1992), so the color
of 1938+666(OIR) is consistent with a lensing galaxy or a lensed radio galaxy
at $z > 1$.
Spiral galaxy $(r-H)$ colors are not widely published, but an
examination of de Jong's work suggests that $2.5 \la (r-H) \la 3$ is
typical while occasional exceptions (e.g. UGC 4256) might be a
magnitude redder. (de Jong, 1995).

Taken together, the close positional coincidence and unusually red
color are strong circumstantial evidence that 1938+666(OIR) is indeed
the gravitational lens system observed in the radio by Patnaik et al.
(1992).  This interpretation might be confirmed by obtaining an image
in sub-arcsecond seeing and comparing the near-IR and radio
morphologies; or (preferably) by obtaining a spectrum, and measuring a
redshift if the object is indeed a quasar.  Either of these tests
might determine if the observed infrared source is the lensed object
or a high redshift lensing galaxy.


The optical-IR colors of many lensed quasars are found to be as red or redder
than observed for 1938+666(OIR) (Lawrence et al.  1994, Larkin et al. 
1994, Annis \& Luppino 1993, Annis 1992).  The most extreme red colors
are seen for MG 0414+0534: $(r-H)=7.8$, and for MG1131 +0456: $(J-K) > 4.1$. 
To compare the reddening directly one needs the redshift of the lens and
the dust responsible for the reddening.  Optical-IR colors of radio
selected quasars can also be quite red.  Webster et al.  1995 report
$B_J - K \approx 2-7$ for the confirmed quasars among the flat-spectrum
radio sources observed with the Parkes telescope, and up to $B_J - K
\approx 8$ for the sources for whom the redshift is not known.  The red
color of 1938+666(OIR) support its identification as a counterpart to the
gravitationally lensed radio ring.

\section{Conclusions}

We have detected an infrared object at the location of the
radio-detected gravitational lens 1938+666, with very red colors
($r-K^{\prime}=6.7 \mag$).  The most natural explanation for
these observations is that the we are seeing the lensed object or a
lensing galaxy at large redshift ($z \ga 1$).  The close positional
coincidence observed is quite unlikely to happen for random field stars
or galaxies, with the exception of the lensing galaxy.  Moreover,
empirical measurements show that $r-K^{\prime}$ colors of most stars and
low to moderate redshift field galaxies are bluer than our observed
limit.  On the other hand, a moderate fraction of flat-spectrum radio
sources and gravitationally lensed quasars are similarly red.  Our
interpretation could be tested by imaging 1938+666 with higher spatial
resolution, or taking a spectrum of this object.

We wish to thank Edwin L. Turner, James E. Gunn, Michael Strauss,
and Robert H. Lupton for useful discussions.
This work has been supported in part by NSF Grant AST94-19400.


\begin{figure}
\caption{
Images of the 1938+666 field in all 5 bands: first row, $K'$ and $H$;
second row, $J$ and $r$; third row, $g$ and $H$ (this time with
1938+666(OIR) indicated by a cross).  Each image is 90 arcseconds on a
side.
North is up, East is to the left.
}
\end{figure}

\begin{table}
\caption{Characteristics of the observations.}
\end{table}

\begin{table}
\begin{tabular}{lccccc}
\cline
Filter & central wave- & bandpass & On source & Individual & Magnitude\\
       &length ($\micron$) & FWHM ($\micron$)  &integration time&
exposure time& \\
\cline 
$K'$ &  2.12 & 0.35 & 12.3 min & 9 s & $17.1 \pm 0.1$ \\
$H$  &  1.65 & 0.30 & 20 min & 25 s & $17.6 \pm 0.1$ \\
$J$  &  1.25 & 0.30 & 12.1 min & 25 s & $19.7 \pm .13$ \\
$r$  & 0.655 & 0.09 & 48 min & 3, 5 min & $23.9 \pm 0.22$\\
$g$  & 0.493 & 0.07 & 48 min & 3, 5 min & $>24.3 (95\%)$ \\
\cline
\end{tabular}
\end{table}

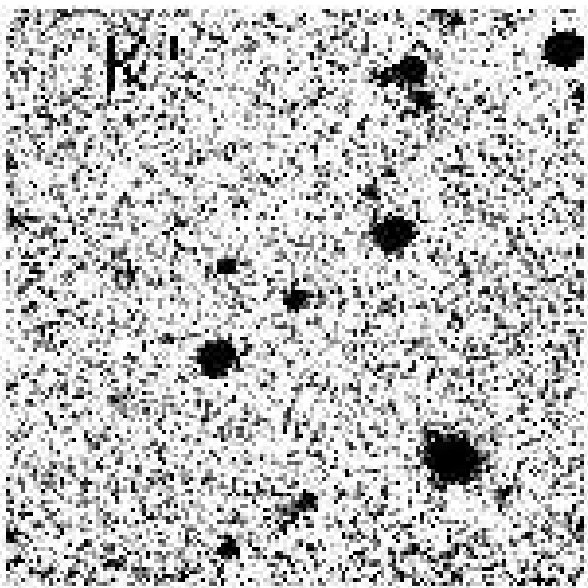
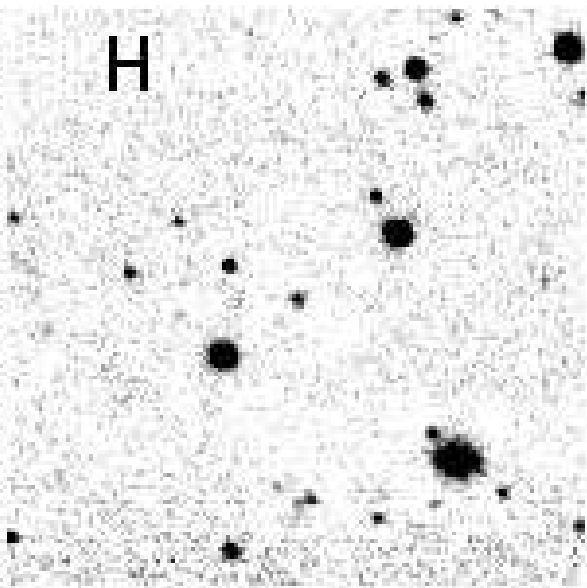
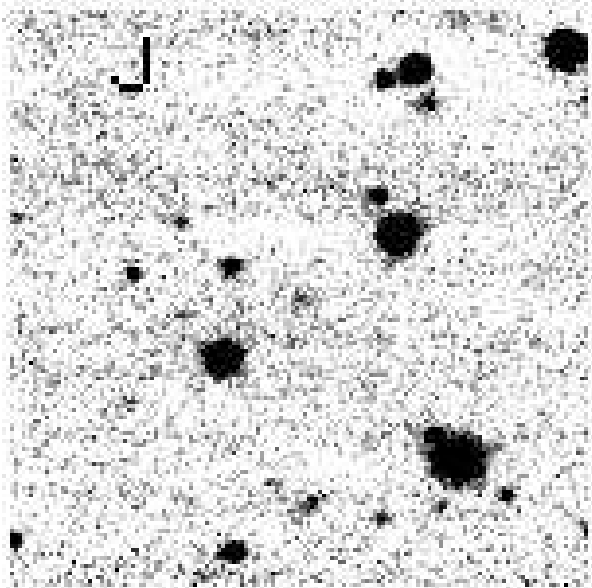
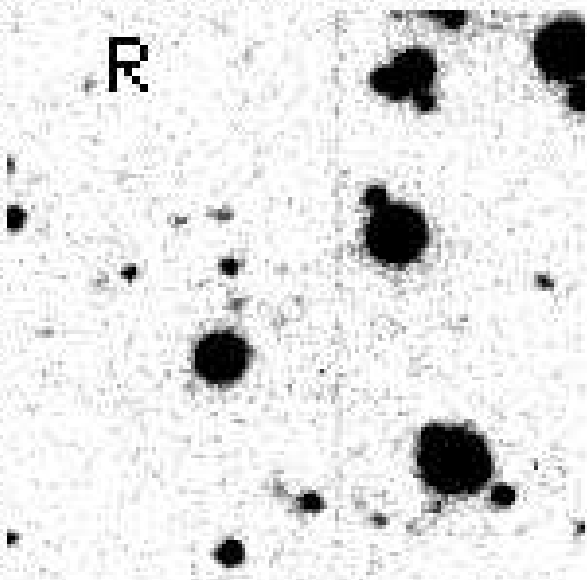
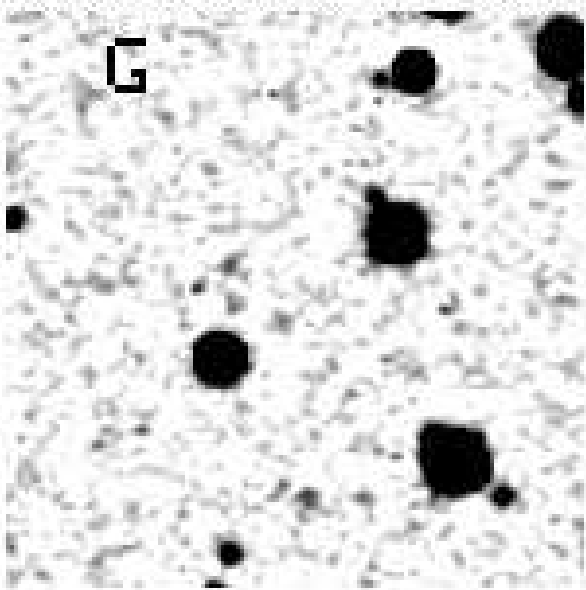
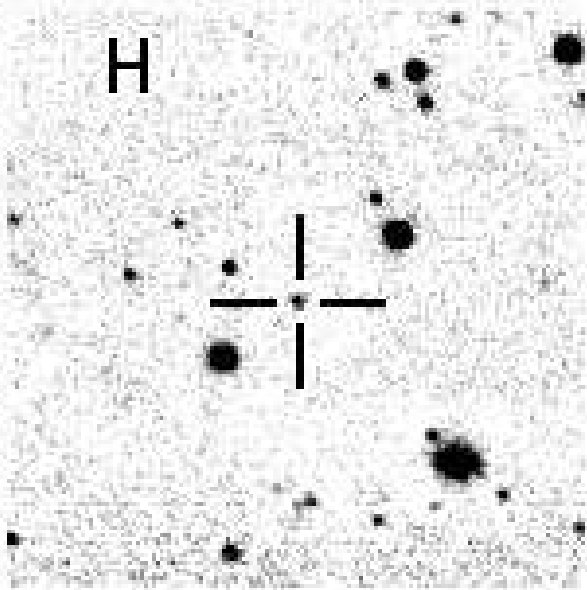
\end{document}